\documentclass[twocolumn,showkeys,aps,prb,showpacs]{revtex4-1}
\usepackage{graphicx}
\usepackage[CJKbookmarks,dvipdfm,colorlinks,linkcolor=blue,citecolor=blue]{hyperref}

\begin{document}

\title{Phonon transport in  $\mathrm{Na_2He}$ at high pressure from a first-principles study}

\author{San-Dong Guo and Ai-Xia Zhang}
\affiliation{$^1$School of Physics, China University of Mining and
Technology, Xuzhou 221116, Jiangsu, China}

\begin{abstract}
Phonon transport of  recently-fabricated $\mathrm{Na_2He}$ at high pressure is investigated  from
 a combination of  first-principles calculations and the linearized phonon Boltzmann equation within
 the single-mode relaxation time approximation (RTA). The calculated room-temperature lattice thermal
 conductivity is 149.19 $\mathrm{W m^{-1} K^{-1}}$, which is very close to one of Si.
 It is found that low-frequency optical modes comprise  16\% of the lattice thermal conductivity, while
high-frequency optical modes have negligible contribution. The high lattice thermal
 conductivity is due to large group  velocities, small Gr$\mathrm{\ddot{u}}$neisen parameters, and long  phonon lifetimes.
 The size effects on lattice thermal conductivity are considered by cumulative thermal conductivity with respect to
phonon mean free path(MFP). To significantly reduce the
lattice thermal conductivity, the  characteristic length smaller than 100 nm is required, and can  reach a  decrease of 36\%.
These results may be useful to understand thermal transport processes that occur inside giant planets.
\end{abstract}

\pacs{72.15.Jf, 71.20.-b, 71.70.Ej, 79.10.-n ~~~~~~~~~~~~~~~~~~~~~~~~~~~~~~~~~~~Email:guosd@cumt.edu.cn}

\maketitle

\section{Introduction}
Helium is chemically inert,  and it is difficult to form thermodynamically
stable compounds. Recently, a stable high-pressure phase $\mathrm{Na_2He}$ with  a fluorite-type structure
is discovered in experiment, which is stable from 113 GPa up to at least 1000 GPa\cite{q1}. By the first-principles calculations, it is predicted that $\mathrm{Na_2He}$ displays  insulating properties, and the energy band gap increases with increasing  pressure. It is interesting and necessary to investigate other physical properties of  $\mathrm{Na_2He}$, such as heat transport.
In semiconductors, the lattice part usually carries the majority of heat around room temperature and
higher, while electronic part has negligible contributions to the thermal conductivity.
The  lattice thermal conductivity  is mainly an anharmonic phenomenon.
Recently, the first-principles calculations can predict anharmonic force constants quantitatively, and then
provide accurate information about the intrinsic phonon-phonon scattering based on the solution
of the phonon Boltzmann transport equation\cite{q2,q3,q4,pv4,q5,q6,q7,q8,q9,q10}. Further, the lattice thermal conductivity can be well reproduced,  being in agreement with experimental results using no adjustable parameters.

Here, we investigate  phonon transport of  $\mathrm{Na_2He}$ with the single-mode RTA  of the linearized phonon
Boltzmann equation. The  lattice thermal conductivity  with respect to temperature is calculated, and the room-temperature value is 149.19 $\mathrm{W m^{-1} K^{-1}}$, which is very close to 155 $\mathrm{W m^{-1} K^{-1}}$ of Si\cite{q11}.
The small Gr$\mathrm{\ddot{u}}$neisen parameters of $\mathrm{Na_2He}$ indicates  a weak  anharmonicity, leading to
the high  thermal conductivity.  The large group  velocities  and long  phonon lifetimes can also  explain high  thermal conductivity.  Moreover, we find that the low-frequency optical modes contribute observably to the total thermal conductivity, while high-frequency optical phonons  can be neglected. To measure phonon MFP,
 the thermal conductivity spectroscopy technique  has been developed\cite{q12}, which can measure MFP distributions over
a wide range of length scales. Therefore, the cumulative lattice thermal conductivity  with respect to phonon MFP is calculated, which can be used to study size effects in heat conduction.
\begin{figure}
  \includegraphics[width=5.5cm]{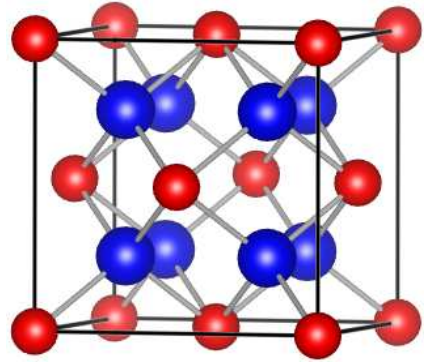}
  \caption{An illustration of  $\mathrm{Na_2He}$ crystal structure. The large blue balls represent Na atoms, and small red balls for He atoms.}\label{st}
\end{figure}

The rest of the paper is organized as follows. In the next
section, we shall give our computational details. In the third section, we shall present phonon transport of $\mathrm{Na_2He}$. Finally, we shall give our discussions and conclusions in the fourth section.

\section{Computational detail}
The  lattice thermal conductivity of $\mathrm{Na_2He}$  is performed with the single mode RTA and linearized phonon Boltzmann equation,  which can be achieved by using Phono3py+VASP codes\cite{pv1,pv2,pv3,pv4}.
For the first-principles calculations, the framework of the all-electron projector augmented wave (PAW) method within the
 density functional theory\cite{1} is employed, as implemented in the package VASP\cite{pv1,pv2,pv3}.
The generalized gradient approximation (GGA)
of Perdew-Burke-Ernzerhof (PBE) is adopted as the exchange-correlation functional\cite{pbe}. A
plane-wave basis set is used  with kinetic energy cutoff of 1000 eV.
The second order harmonic and third
order anharmonic interatomic force constants (IFC)  are
calculated by using a  3 $\times$ 3 $\times$ 3  supercell  and a  2 $\times$ 2 $\times$ 2 supercell, respectively.
Using the harmonic IFCs, the  group velocity  and specific heat can be attained by phonon dispersion relation, and  phonon dispersion  determines the allowed three-phonon scattering processes.
The third-order anharmonic IFCs  can  determine  the phonon lifetimes, which can be attained by calculating the three-phonon scattering rate. To compute lattice thermal conductivities, the
reciprocal spaces of the primitive cells  are sampled using the 20 $\times$ 20 $\times$ 20 meshes.

\begin{figure}
  \includegraphics[width=8cm]{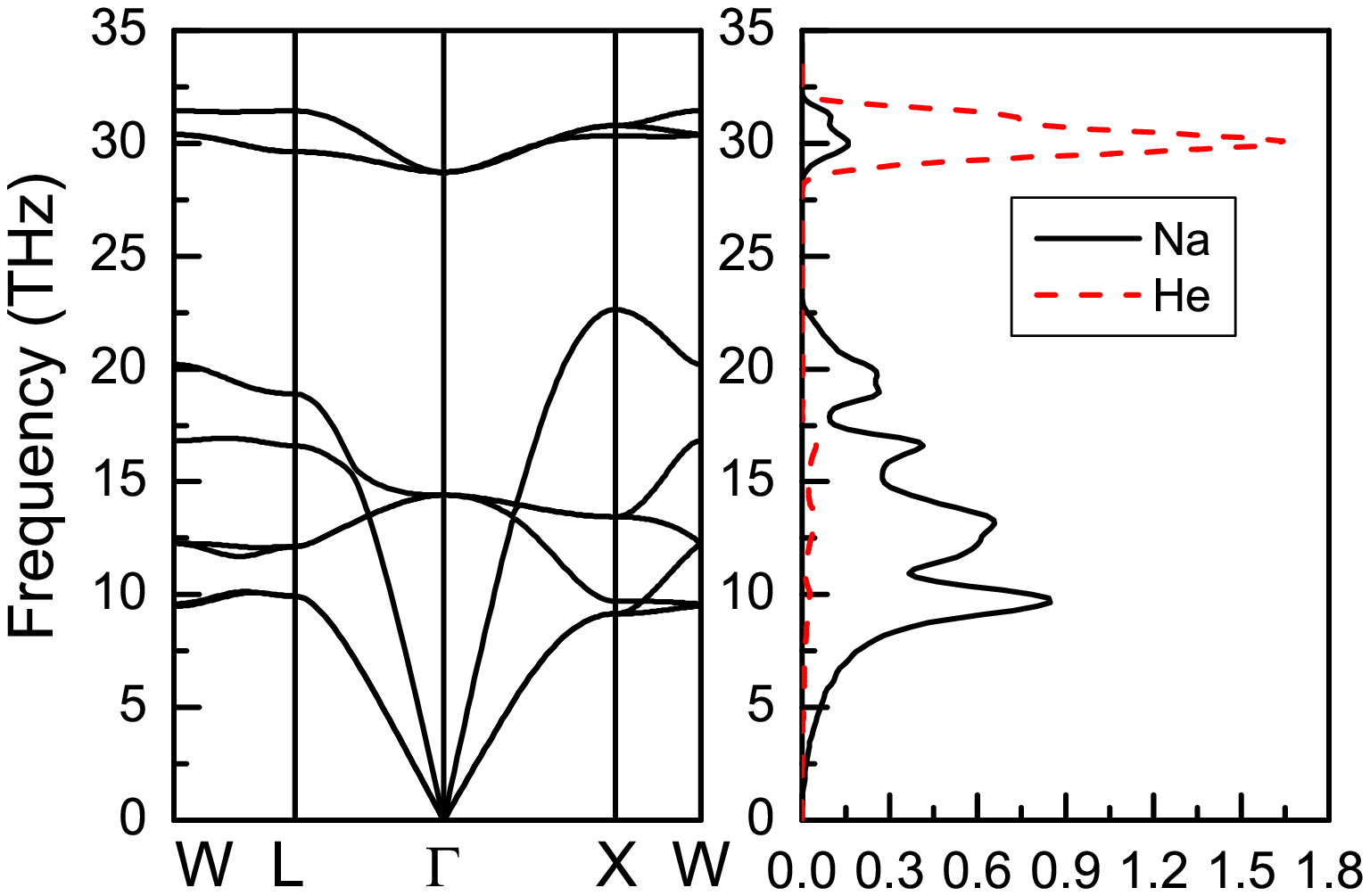}
  \caption{Phonon band structures and partial DOS of  $\mathrm{Na_2He}$.}\label{ph}
\end{figure}
\begin{figure}[!htb]
  \includegraphics[width=7cm]{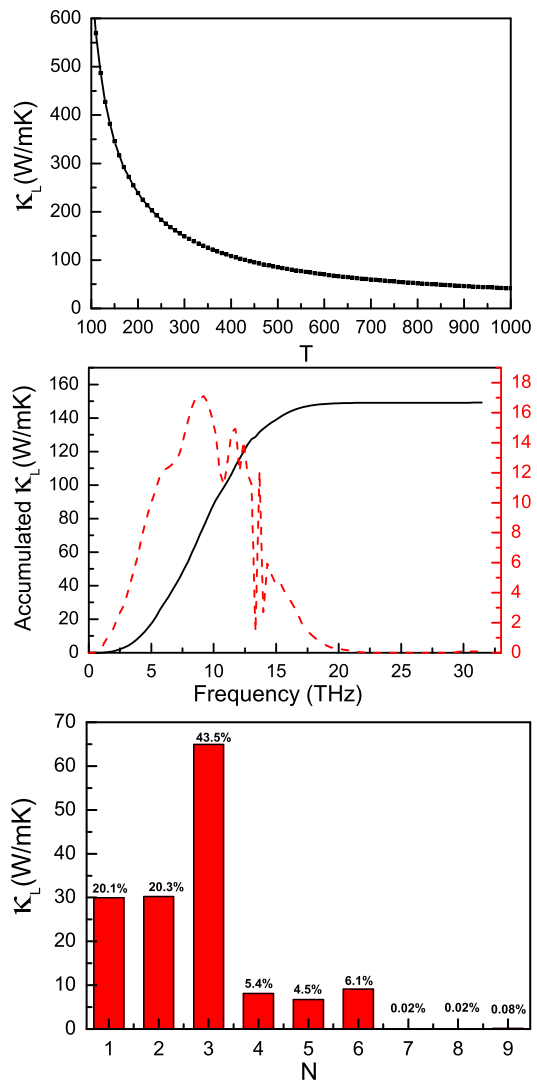}
  \caption{(Color online) Top: the lattice thermal conductivity  of  $\mathrm{Na_2He}$ as a function of temperature; Middle: the accumulated lattice thermal conductivity (300 K), and the derivatives; Bottom: phonon modes contributions toward total lattice thermal conductivity (300 K).}\label{kl}
\end{figure}

\section{MAIN CALCULATED RESULTS AND ANALYSIS}
The $\mathrm{Na_2He}$ has a fluorite-type structure with space
group $Fm\bar{3}m$ at 300 GPa\cite{q1}. The Na and He atoms occupy  the Wyckoff position 8c (0.25,0.25,0.25) and  4a (0,0,0), respectively. The schematic crystal structure is shown in \autoref{st}.
The  experimental lattice parameter a=3.95 $\mathrm{{\AA}}$ is employed to investigate it's phonon transport.
Based on harmonic IFCs, the phonon dispersions are calculated  along  high-symmetry pathes, which  is shown in \autoref{ph}, together with partial density of states (DOS). The unit cell of $\mathrm{Na_2He}$ contains three  atoms, resulting in 3 acoustic and 6 optical phonon branches in the phonon spectra. According to \autoref{ph},  a gap of 6.09 THz can be observed, which separates  three high-frequency optical modes from the low-frequency modes. The high-frequency optical modes  are mainly
from He vibrations, while the acoustic and low-frequency optical modes of phonon dispersions mainly is due to the vibrations of the  Na atoms. The low-frequency optical modes have larger dispersion than high-frequency ones, indicating that low-frequency ones have relatively large group velocities.

\begin{figure}
  \includegraphics[width=7cm]{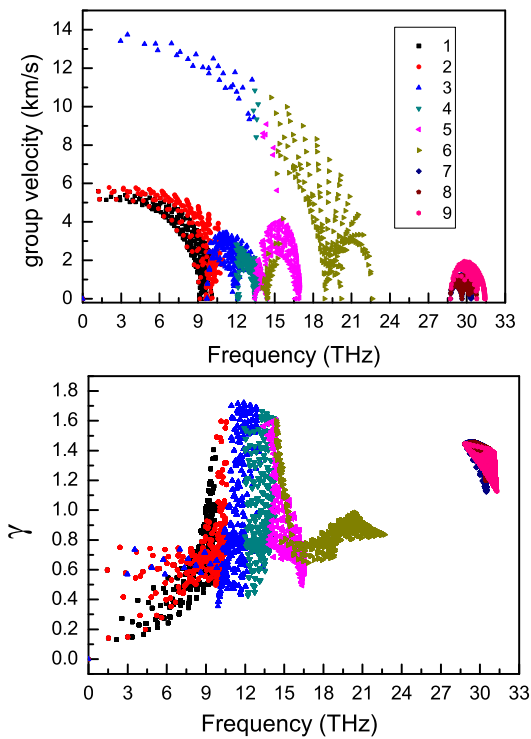}
  \caption{(Color online) The phonon mode group velocities (Top) and mode Gr$\mathrm{\ddot{u}}$neisen parameters (Bottom) of  $\mathrm{Na_2He}$ in the first Brillouin zone.}\label{vg}
\end{figure}
\autoref{kl} shows the lattice thermal conductivity  of  $\mathrm{Na_2He}$ as a function of temperature, the accumulated lattice thermal conductivity (300 K)  along with the derivatives,  and phonon modes contributions toward total lattice thermal conductivity (300 K).
The room-temperature lattice thermal conductivity is 149.19 $\mathrm{W m^{-1} K^{-1}}$, and the lattice thermal conductivity  nearly meets  the $~1/T$ relation at high temperature (above room-temperature). As shown in \autoref{kl}(Middle), the acoustic phonon branches below the 13.34 THz  dominate  lattice thermal conductivity, and   the peak of the derivatives drops  rapidly at 13.34 THz.  The remaining contribution is almost  from low-frequency optical modes, while  the high-frequency optical modes show a neglectful contribution to lattice thermal conductivity,  also found in other semiconductors\cite{q11}. Further, the relative contributions of every acoustic and optical phonon mode to the total lattice thermal conductivity are examined.   The two transverse acoustic (TA) branches have almost the same contributions of about 20\%, while the the longitudinal
acoustic (LA) branch has the largest contributions, as high as 44\%, which is  twice the contribution  of  TA branch.  The low-frequency optical branches provide a contribution of about 16\%, while high -frequency ones only  0.12\%.

The phonon mode group velocities and mode Gr$\mathrm{\ddot{u}}$neisen parameters ($\gamma$)  of  $\mathrm{Na_2He}$ in the first Brillouin zone are plotted in \autoref{vg}. It is found that  the  group velocity of LA branch  is larger  than  ones of  TA branches at low-frequency region, and the largest  group velocity  for LA and TA branches near $\Gamma$ point is 13.74 $\mathrm{km s^{-1}}$ and 5.79  $\mathrm{km s^{-1}}$, respectively.  It is also clearly seen that the third optical branch shows relatively large group velocities due to larger dispersion. Mode Gr$\mathrm{\ddot{u}}$neisen parameters are calculated from third
order anharmonic IFCs, and they all are positive throughout the Brillouin zone.  The mode Gr$\mathrm{\ddot{u}}$neisen parameters
can reflect the strength of anharmonic interactions, and larger  $\gamma$ induces lower  lattice thermal conductivity due to strong anharmonicity.  The high value region of mode Gr$\mathrm{\ddot{u}}$neisen parameters focuses  primarily between 8 THz and 16 THz, and the maxima   is 1.72. The  mode Gr$\mathrm{\ddot{u}}$neisen parameters of $\mathrm{Na_2He}$ are smaller than ones of
BiCuOSe\cite{x1} and PbTe\cite{x2} with low high lattice thermal conductivities as representative thermoelectric materials.
The average mode Gr$\mathrm{\ddot{u}}$neisen parameters is 1.02,  which  means  weak anharmonicity, leading to high lattice thermal conductivity.
\begin{figure}
  \includegraphics[width=7cm]{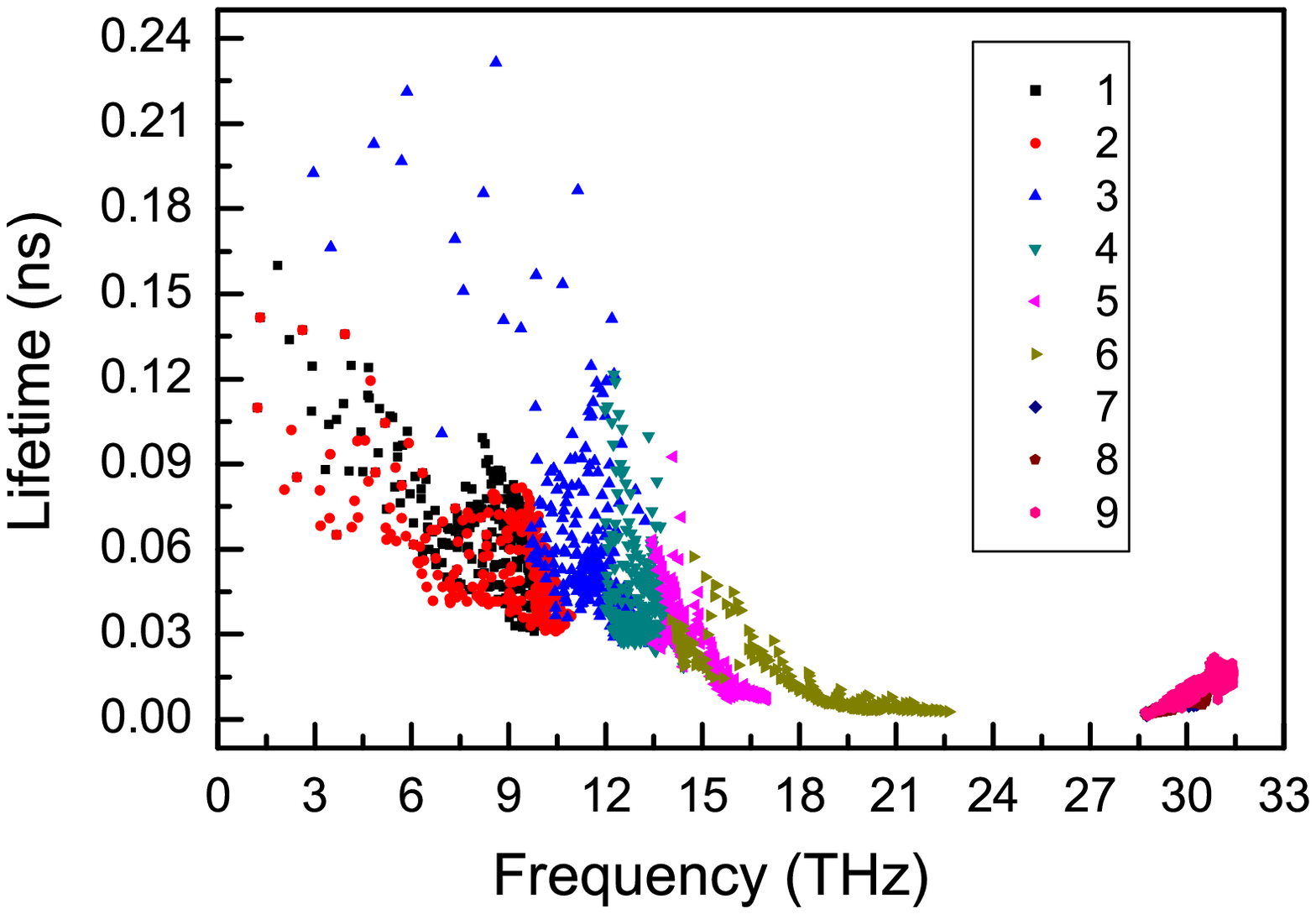}
  \caption{Phonon lifetimes of  $\mathrm{Na_2He}$ at room temperature.}\label{ft}
\end{figure}

The three-phonon scattering rate can be attained by third-order anharmonic IFCs, and
phonon lifetimes can be calculated.
Phonon lifetimes of  $\mathrm{Na_2He}$ at room temperature are shown in \autoref{ft}, which are reciprocal of the phonon linewidth. The lattice thermal conductivity and phonon lifetimes are merely proportional to each other in the single-mode relaxation time method\cite{pv4}. The lifetimes of most acoustic modes  are between 40 ps and 90 ps, and the ones of low-frequency optical modes  decrease with increasing frequency. The first optical branch of they  has long lifetimes,  being well-matched with ones of  acoustic modes.   They lie in 2 ps to 22 ps  for high-frequency optical modes. The distribution of phonon lifetimes can mainly explain the relative contributions of every acoustic and optical phonon mode to the total lattice thermal conductivity

\begin{figure}
  \includegraphics[width=7.0cm]{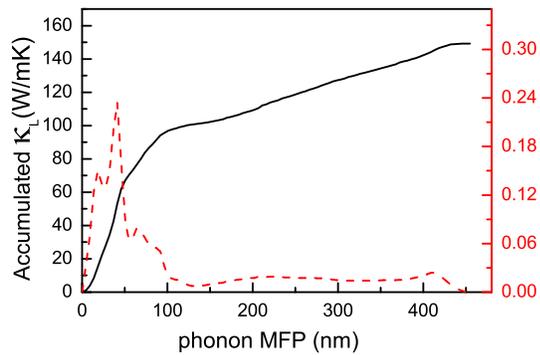}
  \caption{(Color online) Cumulative lattice thermal conductivity of $\mathrm{Na_2He}$ with respect to phonon mean
free path at room temperature.}\label{mkl}
\end{figure}

The cumulative lattice thermal conductivity along with the derivatives  with respect to phonon MFP is shown in \autoref{mkl} at room temperature, which provides information about
the contributions of phonons with different MFP to  the total thermal conductivity. The phonon MFP is also useful to understand and engineer  size effects on lattice thermal conductivity. The total
accumulation  increases with  MFP increasing, and  the accumulation  gradually approaches plateau after MFP reaches 440 nm.
Phonons with MFP smaller than 100
nm contribute  around 64\% to the lattice thermal conductivity.

\section{Discussions and Conclusion}
It is interesting to compare phonon transport of $\mathrm{Na_2He}$ with other semiconductor materials, such as Si and PbTe.
The frequency range of acoustic modes between $\mathrm{Na_2He}$ and Si\cite{q11} is almost the same, and  their phonon lifetimes have the same order of magnitude\cite{q5}. Therefore, the the room-temperature lattice thermal conductivity of $\mathrm{Na_2He}$  (149.19 $\mathrm{W m^{-1} K^{-1}}$)  is very close to 155 $\mathrm{W m^{-1} K^{-1}}$ of Si.
The acoustic frequency range of $\mathrm{Na_2He}$ is very wider than one of PbTe\cite{q6}, which leads to larger group velocities. The  phonon lifetimes of PbTe are very shorter than ones of $\mathrm{Na_2He}$.
So, the lattice thermal conductivity (1.9 $\mathrm{W m^{-1} K^{-1}}$) of  PbTe at 300 K is very lower than one of  $\mathrm{Na_2He}$. The similarity  between $\mathrm{Na_2He}$ and PbTe is that the contributions
of optical phonons to lattice thermal conductivity  remain very large, 16\% for $\mathrm{Na_2He}$ and 22\% for PbTe
at room temperature.  However, for Si,  the optical branches  comprise
only 5\% of the lattice thermal conductivity\cite{f1,f2,f3}.  The difference  among  them is  size effects on lattice thermal conductivity, which can be described by cumulative lattice thermal conductivity as a function of MFP.
 MFPs smaller than 10
nm contribute to around 90\% of the lattice thermal conductivity for PbTe\cite{q6}. Phonons with MFP smaller than 100
nm comprise  around 64\% of  the lattice thermal conductivity for $\mathrm{Na_2He}$.
MFPs smaller than 1000 nm contribute to almost half of the total thermal conductivity for Si\cite{q5}.

In summary,  the intrinsic lattice thermal
conductivity of $\mathrm{Na_2He}$ is calculated  based mainly on the reliable first-principle calculations and Boltzmann transport theory. The room-temperature lattice  thermal conductivity is found to be
149.19 $\mathrm{W m^{-1} K^{-1}}$. The acoustic and low-frequency optical branches provide nearly 100\%
contribution  to total lattice thermal conductivity.  The high lattice thermal conductivity is due to weak anharmonicity and high group velocities. In addition, the size effects on  thermal conductivity are also studied by cumulative lattice thermal conductivity  with respect to MFP. The present
work can  encourage further  efforts to investigate other chemical and physical properties of $\mathrm{Na_2He}$.

\begin{acknowledgments}
This work is supported by the National Natural Science Foundation of China (Grant No. 11404391). We are grateful to the Advanced Analysis and Computation Center of CUMT for the award of CPU hours to accomplish this work.
\end{acknowledgments}


\begin{references}
\bibitem{q1}D. Xiao et al., Nat. Chem.  (2017).\\
http://dx.doi.org/10.1038/nchem.2716


\bibitem{q2} L. Chaput, Phys. Rev. Lett. \textbf{110}, 265506 (2013).

\bibitem{q3} A. Ward, D. A. Broido, D. A. Stewart and G. Deinzer, Phys.
Rev. B \textbf{80}, 125203 (2009).

\bibitem{q4} A. Ward and D. A. Broido, Phys. Rev. B \textbf{81}, 085205
(2010).

\bibitem{pv4}A. Togo, L. Chaput and I. Tanaka, Phys. Rev. B \textbf{91}, 094306 (2015).

\bibitem{q5} K. Esfarjani,G. Chen and H. T. Stokes, Phys. Rev. B \textbf{84}, 085204
(2011).

\bibitem{q6} Z. Tian, J. Garg, K. Esfarjani, T. Shiga, J. Shiomi and G. Chen,
Phys. Rev. B \textbf{85}, 184303 (2012).

\bibitem{q7} T. Shiga, J. Shiomi, J. Ma, O. Delaire, T. Radzynski, A.
Lusakowski,K. Esfarjani and G. Chen,Phys. Rev. B \textbf{85}, 155203
(2012).

\bibitem{q8} W. Li, J. Carrete, N. A Katcho and N. Mingo, Comput. Phys.
Commun. \textbf{185}, 1747 (2014).

\bibitem{q9} L. Lindsay, D. A. Broido and T. L. Reinecke, Phys. Rev. Lett.
\textbf{109}, 095901 (2012).

\bibitem{q10}L. Lindsay, D. A. Broido and T. L. Reinecke, Phys. Rev. Lett.
\textbf{111}, 025901 (2013).


\bibitem{q11}L. Lindsay, D. A. Broido and T. L. Reinecke, Phys. Rev. B \textbf{87}, 165201 (2013).


\bibitem{q12}A. J. Minnich, J. A. Johnson, A. J. Schmidt, K. Esfarjani, M. S. Dresselhaus, K. A. Nelson and G. Chen,
Phys. Rev. Lett. \textbf{107}, 095901 (2011).
\bibitem{1}P. Hohenberg and W. Kohn, Phys. Rev. \textbf{136},
B864 (1964); W. Kohn and L. J. Sham, Phys. Rev. \textbf{140},
A1133 (1965).





\bibitem{pv1} G. Kresse, J. Non-Cryst. Solids \textbf{193}, 222 (1995).

\bibitem{pv2} G. Kresse and J. Furthm$\ddot{u}$ller, Comput. Mater. Sci. 6, \textbf{15} (1996).

\bibitem{pv3} G. Kresse and D. Joubert, Phys. Rev. B \textbf{59}, 1758 (1999).

\bibitem{pbe}J. P. Perdew, K. Burke and M. Ernzerhof, Phys. Rev. Lett. \textbf{77}, 3865 (1996).



\bibitem{x1}H. Z. Shao, X. J. Tan, G. Q. Liu, J. Jiang and  H. C. Jiang, Sci.
Rep. \textbf{6}, 21035 (2016).

\bibitem{x2}Y. Zhang, X. Z.  Ke, C. F. Chen, J. Yang and P. R. C. Kent,  Phys.
Rev. B \textbf{80}, 024304 (2009).



\bibitem{f1}A. S. Henry and G. Chen, J. Comput. Theor. Nanosci. \textbf{5}, 141 (2008).

\bibitem{f2}D. A. Broido, M. Malorny, G. Birner, N.Mingo, and D. A. Stewart,
Appl. Phys. Lett. \textbf{91}, 231922 (2007).

\bibitem{f3}D. P. Sellan, J. E. Turney, A. J. H. McGaughey, and C. H. Amon, J
Appl. Phys. \textbf{108}, 113524 (2010).



\end{references}
\end{document}